\documentclass[12pt]{article}

\usepackage{latexsym}
\usepackage{amssymb,amsfonts,amsmath}
\usepackage{graphicx} 
\usepackage{indentfirst}
\usepackage{bbm}
\usepackage{amssymb}
\usepackage{verbatim}
\usepackage{amsmath, amsthm,amssymb}
\usepackage{mathrsfs}
\usepackage{hyperref}
\usepackage{amsfonts}
\usepackage{dsfont}
\usepackage{cite}
\usepackage{xcolor}

\parskip=\medskipamount

\newcommand {\cD}{{\cal D}}

\newcommand {\cN}{{\cal N}}

\def\a{\alpha}

\def\b{\beta}

\def\d{\delta}

\def\g{\gamma}

\def\j{\psi}

\def\m{\mu}

\def\U{\Upsilon}

\newcommand{\ad}{{\dot{\alpha}}}                           
\newcommand{\bd}{{\dot{\beta}}}                            
\newcommand{\ve}{\varepsilon}                            

\newcommand{\hf}{\frac12}

\newcommand{\be}{\begin{equation}}
\newcommand{\ee}{\end{equation}}
\newcommand{\bea}{\begin{eqnarray}}
\newcommand{\eea}{\end{eqnarray}}

\newcommand{\bm}[1]{\mbox{\boldmath$#1$}}

\def\double #1{#1{\hbox{\kern-2pt $#1$}}}

\newcommand{\gd}{{\dot\g}}
\newcommand{\dd}{{\dot\d}}

\newif\ifdtup

\newcommand{\bsubeq}{\begin{subequations}}
\newcommand{\esubeq}{\end{subequations}}

\numberwithin{equation}{section}

\begin{document}

\begin{titlepage}
\begin{flushright}
October, 2019\\
\end{flushright}
\vspace{5mm}

\begin{center}
{\Large \bf Spin projection operators in (A)dS
and partial masslessness}\\ 
\end{center}

\begin{center}

{\bf   
Sergei M. Kuzenko and Michael Ponds 
} \\
\vspace{5mm}

\footnotesize{
{\it Department of Physics M013, The University of Western Australia\\
35 Stirling Highway, Perth W.A. 6009, Australia}}  
\vspace{2mm}
~\\
Email: \texttt{ 
sergei.kuzenko@uwa.edu.au, michael.ponds@research.uwa.edu.au}\\
\vspace{2mm}

\end{center}

\begin{abstract}
\baselineskip=14pt
We elaborate on the traceless and transverse spin projectors in four-dimensional de Sitter and anti-de Sitter spaces. The poles of these projectors are shown to correspond to partially massless fields. We also obtain a factorisation of the conformal operators 
associated with gauge fields of arbitrary Lorentz type 
$(m/2,n/2 )$, with $m$ and $n$  positive integers. 
\end{abstract}

\vfill

\vfill
\end{titlepage}

\newpage
\renewcommand{\thefootnote}{\arabic{footnote}}
\setcounter{footnote}{0}

%

\allowdisplaybreaks


\section{Introduction}

In four-dimensional Minkowski space ${\mathbb M}^4$, spin projection operators, 
also known as traceless and  transverse (TT) spin-$s$ projectors, were constructed
by Behrends and Fronsdal more than sixty years ago
\cite{BF,Fronsdal}. These TT projectors have found numerous applications. 
For instance, it is well-known that they determine the structure of 
massive spin-$s$ propagators in the quantum theory. 
They can also be used to construct gauge-invariant actions.
An important example of the latter application is the formulation of conformal 
higher-spin actions proposed by Fradkin and Tseytlin \cite{FT}, although 
the TT spin-$s$ projectors were given  explicitly
in \cite{FT} only for $s\leq 2$.

Refs. \cite{BF,Fronsdal} made use of the four-vector notation in conjunction with the four-component spinor formalism, which resulted in rather complicated expressions for the TT spin-$s$ projectors.
However, switching to the two-component spinor formalism leads to 
remarkably simple and compact expressions for these projectors \cite{SG,GGRS}.

For higher-dimensional Minkowski space ${\mathbb M}^d$, with $d>4$, the 
TT
integer-spin projectors were constructed by Segal \cite{Segal} 
who used these  operators to formulate bosonic conformal higher-spin actions. 
In the literature, there have appeared 
different forms of the bosonic TT projectors \cite{FMS,PT,Bonezzi,IP}, 
which
may be shown to be equivalent to the ones presented in the arXiv version
of \cite{Segal}. 
The TT half-integer-spin projectors for $d>4$ were constructed for the first time 
in \cite{IP}. It should be pointed out that 
the three-dimensional case is somewhat special, and the corresponding 
spin projection operators were described in \cite{BKLFP} (see also \cite{BHHK} for the superprojectors).

Unlike Minkowski space,  both de Sitter (dS) and anti-de Sitter (AdS) spaces 
have non-vanishing curvature, which makes it more challenging to construct 
the TT spin-$s$ projectors.
For this reason only the lower-spin cases corresponding to $s\leq 2$ have been considered in the literature \cite{DW2}. In this paper we construct all 
the spin projection operators in (A)dS${}_4$
and use them to  derive various properties of higher-spin (i) 
partially massless fields \cite{DW2, DeserN1, DeserN2, DeserW1, DeserW3, DeserW4, Higuchi1, Higuchi2, Higuchi3, Zinoviev,Metsaev}; (ii) massive  fields;
and (iii) conformal  models.

Throughout the paper we make use of the two-component spinor formalism and follow
the notation and conventions of \cite{BK}.
 In this notation
 the algebra of AdS covariant derivatives\footnote{In vector notation this reads $\big[\mathcal{D}_a,\mathcal{D}_b\big]=-2\mathcal{S}^2M_{ab}$.}
is
 \begin{align}
 \big[\mathcal{D}_{\a\ad},\mathcal{D}_{\b\bd}\big]=-4\mathcal{S}^2\big(\ve_{\a\b}\bar{M}_{\ad\bd}+\ve_{\ad\bd}M_{\a\b}\big)~, \label{2.1}
 \end{align} 
 where the Lorentz generator $M_{\a\b}$ is defined by $M_{\a\b} \j_\g = 
 \ve_{\g(\a} \j_{\b)}$.
 An analysis similar to that given below applies in the case of de Sitter space, one just needs to replace all occurrences of $\mathcal{S}^2$ with $-\mathcal{S}^2$.


\section{Spin  projection operators}

 Of crucial importance to our subsequent analysis  is  the  quadratic AdS Casimir operator\footnote{This operator
may be compared with the quadratic Casimir operator of 
the $\cN=1$ AdS supergroup \cite{BKS}.}
 \begin{align}
\mathcal{Q}:=\Box-2\mathcal{S}^2(M^{\g\d}M_{\g\d}+\bar{M}^{\gd\dd}\bar{M}_{\gd\dd})~,\qquad \big[\mathcal{Q},\mathcal{D}_{\a\ad}\big]=0~,
\label{2.2}
\end{align}
where $\Box = \cD^a \cD_a = - \hf \cD^{\a\ad} \cD_{\a\ad}$. 
 
Denote by $V_{(m,n)}$ the space of fields $\phi_{\a(m)\ad(n)}$ which are totally symmetric in their dotted and, independently, in their undotted indices.\footnote{This means that from 
the beginning we work with traceless tensor fields.} 
 To construct the projectors, we introduce two operators $\mathbb{P}^{(m,n)},~\widehat{\mathbb{P}}^{(m,n)}:V_{(m,n)}\rightarrow V_{(m,n)}$ 
 which are 
defined by their action on $\phi_{\a(m)\ad(n)}$ as
\begin{subequations}\label{2.3}
\begin{align}
\mathbb{P}_{\a(m)\ad(n)}(\phi)&=\mathcal{D}_{(\ad_1}{}^{\b_1}\cdots\mathcal{D}_{\ad_n)}{}^{\b_n}\mathcal{D}_{(\b_1}{}^{\bd_1}\cdots\mathcal{D}_{\b_n}{}^{\bd_n}\phi_{\a_1\dots\a_m)\bd(n)}~, \label{2.3a}\\[6pt]
\widehat{\mathbb{P}}_{\a(m)\ad(n)}(\phi)&=\mathcal{D}_{(\a_1}{}^{\bd_1}\cdots\mathcal{D}_{\a_m)}{}^{\bd_m}\mathcal{D}_{(\bd_1}{}^{\b_1}\cdots\mathcal{D}_{\bd_m}{}^{\b_m}\phi_{\b(m)\ad_1\dots\ad_n)}~. \label{2.3b}
\end{align}
\end{subequations}
Both operators \eqref{2.3} project out the transverse component of the field $\phi_{\a(m)\ad(n)}$,
\begin{subequations}\label{2.4}
\begin{align}
\mathcal{D}^{\b\bd}\mathbb{P}_{\b\a(m-1)\bd\ad(n-1)}(\phi)&=0~,\\
\mathcal{D}^{\b\bd}\widehat{\mathbb{P}}_{\b\a(m-1)\bd\ad(n-1)}(\phi)&=0~.
\end{align} 
\end{subequations}
However they are not projectors in the sense that they do not square to themselves. In fact, one may show that they instead satisfy
\begin{subequations}\label{2.5}
\begin{align}
\mathbb{P}^{(m,n)}\mathbb{P}^{(m,n)}\phi_{\a(m)\ad(n)}=\prod_{t=1}^{n}(\mathcal{Q}-\lambda_{(t,m,n)}\mathcal{S}^2)\mathbb{P}^{(m,n)}\phi_{\a(m)\ad(n)}~, \label{2.5a}\\
\widehat{\mathbb{P}}^{(m,n)}\widehat{\mathbb{P}}^{(m,n)}\phi_{\a(m)\ad(n)}=\prod_{t=1}^{m}(\mathcal{Q}-\lambda_{(t,m,n)}\mathcal{S}^2)\widehat{\mathbb{P}}^{(m,n)}\phi_{\a(m)\ad(n)}~, \label{2.5b}
\end{align}
\end{subequations}
where the parameters $\lambda_{(t,m,n)}$ are defined by
\begin{align}
 \lambda_{(t,m,n)}&:= (m+n-t+3)(m+n-t-1)+(t-1)(t+1)~.\label{2.6}
\end{align}
From Eq. \eqref{2.5} it follows that the two operators $\Pi^{(m,n)},~\widehat{\Pi}^{(m,n)}:V_{(m,n)}\rightarrow V_{(m,n)}$ where
\begin{subequations}\label{2.7}
\begin{align}
\Pi_{\a(m)\ad(n)}(\phi)=\bigg[\prod_{t=1}^{n}(\mathcal{Q}-\lambda_{(t,m,n)}\mathcal{S}^2)\bigg]^{-1}\mathbb{P}_{\a(m)\ad(n)}(\phi)~, \label{2.7a}\\
\widehat{\Pi}_{\a(m)\ad(n)}(\phi)=\bigg[\prod_{t=1}^{m}(\mathcal{Q}-\lambda_{(t,m,n)}\mathcal{S}^2)\bigg]^{-1}\widehat{\mathbb{P}}_{\a(m)\ad(n)}(\phi)~, \label{2.7b}
\end{align}
\end{subequations}
 square to themselves and project out the transverse subspace of $V_{(m,n)}$,
 \begin{subequations}\label{2.8}
\begin{align}
\Pi^{(m,n)}\Pi^{(m,n)}\phi_{\a(m)\ad(n)}&=\Pi^{(m,n)}\phi_{\a(m)\ad(n)}~,\qquad \mathcal{D}^{\b\bd}\Pi_{\b\a(m-1)\bd\ad(n-1)}(\phi)=0~,\\
\widehat{\Pi}^{(m,n)}\widehat{\Pi}^{(m,n)}\phi_{\a(m)\ad(n)}&=\widehat{\Pi}^{(m,n)}\phi_{\a(m)\ad(n)}~,\qquad \mathcal{D}^{\b\bd}\widehat{\Pi}_{\b\a(m-1)\bd\ad(n-1)}(\phi)=0~.
\end{align}
\end{subequations}
Therefore the operators \eqref{2.7} are the spin projection operators in AdS. Actually, the two types of projectors prove to coincide,
\begin{align}\label{2.9}
\Pi_{\a(m)\ad(n)}(\phi)=\widehat{\Pi}_{\a(m)\ad(n)}(\phi)~,
\end{align}
and so it suffices to consider only the first, \eqref{2.7a}.

In addition, it is possible to show that for any field $\phi_{\a(m)\ad(n)}$ such that the projector \eqref{2.7a} is well defined, there exists some $\zeta_{\a(m-1)\ad(n-1)}$ such that 
\begin{align}
\big(\mathds{1}-\Pi^{(m,n)}\big)\phi_{\a(m)\ad(n)}=\mathcal{D}_{(\a_1(\ad_1}\zeta_{\a_2\dots\a_{m)\ad_2\dots\ad_n)}}~. \label{2.10}
\end{align}
This means that any field may be decomposed 
as\footnote{A supersymmetric extension of \eqref{2.11} is given in 
\cite{BKS}.}
\begin{align}
\phi_{\a(m)\ad(n)}=\sum_{l=0}^{m-1}\mathcal{D}_{(\a_1(\ad_1}\dots\mathcal{D}_{\a_l\ad_l}\phi^{\text{T}}_{\a_{l+1}\dots\a_{m})\ad_{l+1}\dots\ad_n)}+\mathcal{D}_{(\a_1(\ad_1}\dots\mathcal{D}_{\a_m)\ad_m}\phi_{\ad_{m+1}\dots\ad_{n})}~, \label{2.11}
\end{align}
for some set of fields $\{\phi^{\text{T}}_{\a(m)\ad(n)},\phi^{\text{T}}_{\a(m-1)\ad(n-1)},\dots,\phi^{\text{T}}_{\a\ad(n-m+1)} \}$ which are transverse and where we have assumed, without loss of generality, that $n\geq m$.


\section{Analysis of results and applications}

It is of interest to understand the physical significance of the parameters \eqref{2.6} which appear in the definition of the projectors \eqref{2.7}. With this in mind we now introduce on-shell fields, which are those satisfying the equations
\begin{subequations}\label{eom}
\begin{align}
\big(\mathcal{Q}-\mu^2\big)\phi_{\a(m)\ad(n)}&=~0~,\label{eom1}\\
\mathcal{D}^{\g\gd}\phi_{\a(m-1)\g\ad(n-1)\gd}&=~0~.\label{eom2}
\end{align}
\end{subequations}
We say that such a field describes a spin $s=\frac{1}{2}(m+n)$ particle with pseudo-mass\footnote{This terminology is because with our conventions $\mu$ is not the physical mass. For example, in the massless case $\mu\neq 0$.} $\mu$.  


\subsection{Partially massless fields}

It is typical to choose $m=n=s$ for spin-$s$ bosonic fields whereas the usual choice for spin-$\big(s+\frac 12\big)$ fermionic fields is $m=n-1=s$. By now it is well known that in these cases, the system of equations \eqref{eom} becomes invariant under gauge transformations of depth $t$
\begin{subequations}
\begin{align}
\delta_{\zeta}\phi_{\a(s)\ad(s)}&=\mathcal{D}_{(\a_1(\ad_1}\dots\mathcal{D}_{\a_t\ad_t}\zeta_{\a_{t+1}\dots\a_s)\ad_{t+1}\dots\ad_{s})}~,\\
\delta_{\zeta}\phi_{\a(s)\ad(s+1)}&=\mathcal{D}_{(\a_1(\ad_1}\dots\mathcal{D}_{\a_t\ad_t}\zeta_{\a_{t+1}\dots\a_s)\ad_{t+1}\dots\ad_{s+1})}~,
\end{align} 
\end{subequations}
for an on-shell gauge parameter when the mass squared takes the special values\footnote{Due to our definition \eqref{eom1}, the mass values \eqref{3.3} are shifted with respect to the usual ones.}\cite{DeserW4, Zinoviev, Metsaev}
\begin{subequations}\label{3.3}
\begin{align}
\mu^2_{(t,s)}&=[(2s-t+3)(2s-t-1)+(t-1)(t+1)]\mathcal{S}^2~,\label{3.3a}\\
\mu^2_{(t,s+\frac 12)}&=[(2s-t+4)(2s-t)+(t-1)(t+1)]\mathcal{S}^2~, \label{3.3b}
\end{align}
\end{subequations}
where $1\leq t \leq s$. Strictly massless fields correspond to $t=1$ whilst all other values of $t$ correspond to partially massless fields. Remarkably, we see that for these values of $m$ and $n$, the partially massless values coincide with the parameters in the projectors,
\begin{align}
\mu^2_{(t,s)}=\lambda_{(t,s,s)}\mathcal{S}^2~,\qquad \mu^2_{(t,s+\frac 12)}=\lambda_{(t,s,s+1)}\mathcal{S}^2~.
\end{align} 
Therefore, we can extend this notion and say that a field $\phi_{\a(m)\ad(n)}$ is partially massless when it satisfies \eqref{eom} with 
\begin{align}
\mu^2_{(t,m,n)}=\lambda_{(t,m,n)}\mathcal{S}^2~,\qquad 1\leq t \leq \text{min}(m,n)~.\label{3.5}
\end{align}
Indeed, as shown in the appendix, for these values a gauge invariance with depth $t$ emerges in the system of equations \eqref{eom},
\begin{align}
\delta_{\zeta}\phi_{\a(m)\ad(n)}=\mathcal{D}_{(\a_1(\ad_1}\dots\mathcal{D}_{\a_t\ad_t}\zeta_{\a_{t+1}\dots\a_{m})\ad_{t+1}\dots\ad_n)}~, \label{3.6}
\end{align}
with $\zeta_{\a(m-t)\ad(n-t)}$ being on-shell. 
This gauge symmetry\footnote{Gauge-invariant Lagrangian formulations for partially massless fields in (A)dS were given in 
\cite{Zinoviev, Metsaev,SV}.}
 is our main motivation for choosing the upper bound of min$(m,n)$ for $t$ in the definition \eqref{3.5}.

For transverse fields $\phi^{\text{T}}_{\a(m)\ad(n)}$ satisfying  \eqref{eom2},
it follows from \eqref{2.5} that upon application of $\mathbb{P}^{(m,n)}$ and $\widehat{\mathbb{P}}^{(m,n)}$, we obtain the following factorisations
\begin{subequations}\label{3.10}
\begin{align}
\mathbb{P}_{\a(m)\ad(n)}(\phi^{\text{T}})&=\prod_{t=1}^{n}(\mathcal{Q}-\lambda_{(t,m,n)}\mathcal{S}^2)\phi^{\text{T}}_{\a(m)\ad(n)}~,\label{3.10a}\\
\widehat{\mathbb{P}}_{\a(m)\ad(n)}(\phi^{\text{T}})&=\prod_{t=1}^{m}(\mathcal{Q}-\lambda_{(t,m,n)}\mathcal{S}^2)\phi^{\text{T}}_{\a(m)\ad(n)}~.\label{3.10b}
\end{align} 
\end{subequations}
So partially massless fields may be understood as those for which \eqref{3.10} vanishes, or alternatively as those fields whose masses appear as poles in the projectors \eqref{2.7}.


\subsection{Massive fields}

Massive fields correspond to those values of $\m^2$ which differ from \eqref{3.5}.
 The tensor field 
  $\phi_{\a(m)\ad(n)}$ is not the only type of field that can be used to describe a massive spin $s=\frac{1}{2}(m+n)$ particle in AdS. Such representations can be equivalently realised on $V_{(m+t,n-t)}$ where $1\leq t \leq n$. To see this, consider the set of operators 
\begin{align}
\Delta^{(t,m,n)}_{~\a\ad}&=\frac{\mathcal{D}_{\a\ad}}{\sqrt{\mathcal{Q}-\lambda_{(t,m,n)}\mathcal{S}^2}}~.\label{3.7}
\end{align}
For a fixed $t$ the $\lambda_{(t,m,n)}$ with different values of $m$ and $n$ such that $m+n=\text{const}$ are all equal, therefore we may drop these two labels in \eqref{3.7}, $\Delta^{(t)}\equiv\Delta^{(t,m,n)} $, if it is clear which family of fields it acts upon. If $\phi^{\text{T}}_{\a(m)\ad(n)}$ is transverse \eqref{eom2}, then we may use $\Delta^{(t)}$ to convert back and forth between tensor types,
\begin{subequations}\label{3.8}
\begin{align}
\phi^{\text{T}}_{\a(m+t)\ad(n-t)}&=\Delta^{(n-t+1)\bd_{1}}_{~~~~~~\a_{1}}\cdots\Delta^{(n)\bd_{t}}_{~~\a_{t}}\phi^{\text{T}}_{\a_{t+1}\dots\a_{m+t}\ad(n-t)\bd(t)}~,\label{3.8a}\\
\phi^{\text{T}}_{\a(m)\ad(n)}&=\Delta^{(n)\b_{1}}_{~~\ad_{1}}\cdots \Delta^{(n-t+1)\b_{t}}_{~~~~~~\ad_{t}}\phi^{\text{T}}_{\a(m)\b(t)\ad_{t+1}\dots\ad_n}~. \label{3.8b}
\end{align}
\end{subequations}
Alternatively, we could instead use \eqref{3.7} to trade dotted indices for undotted ones and convert to fields of the type belonging to $V_{(m-t,n+t)}$,
\begin{subequations}\label{3.9}
\begin{align}
\phi^{\text{T}}_{\a(m-t)\ad(n+t)}&=\Delta^{(m-t+1)\b_{1}}_{~~~~~~\ad_{1}}\cdots\Delta^{(m)\b_{t}}_{~~\ad_{t}}\phi^{\text{T}}_{\a(m-t)\b(t)\ad_{t+1}\dots\ad_{n+t}}~,\label{3.9a}\\
\phi^{\text{T}}_{\a(m)\ad(n)}&=\Delta^{(m)\bd_{1}}_{~~\a_{1}}\cdots \Delta^{(m-t+1)\bd_{t}}_{~~~~~~\a_{t}}\phi^{\text{T}}_{\a_{t+1}\dots\a_{m}\ad(n)\bd(t)}~, \label{3.9b}
\end{align}
\end{subequations}
 where now $1\leq t \leq m$. We note that in \eqref{3.8a} and \eqref{3.9a}, each of the fields $\phi^{\text{T}}_{\a(m\pm t)\ad(n\mp t)}$ inherit their transversality from $\phi^{\text{T}}_{\a(m)\ad(n)}$ and so their right hand sides are totally symmetric. 
 
 The denominator of \eqref{3.7} is well defined for all on-shell fields except for those whose mass satisfies \eqref{3.5}. This means that for partially massless fields, the different spaces $V_{(m,n)}$ with $m+n=\text{const}$ describe inequivalent representations, which is not surprising from the point of view of gauge symmetry.  

The operators \eqref{3.7} may also be used to rewrite the projectors \eqref{2.7} 
as\footnote{We would like to point out that the spin projection operators for conformal fields in arbitrary conformally-flat backgrounds appeared in \cite{KP2}. Their structure is similar to \eqref{3.11}. However, the projectors of Ref. \cite{KP2} were formulated using the conformal covariant derivative and it is non-trivial  to switch to a description in terms of the torsion-free Lorentz covariant derivative used here.}  
\begin{subequations}\label{3.11}
\begin{align}
\Pi_{\a(m)\ad(n)}(\phi)&=\Delta^{(n)\b_n}_{~(\ad_{n}}\cdots \Delta^{(1)\b_1}_{~\ad_{1})}\Delta^{(1)\bd_1}_{~(\b_{1}}\cdots \Delta^{(n)\bd_n}_{~\b_{n}}\phi_{\a_1\dots\a_m)\bd(n)}~,\\
\widehat{\Pi}_{\a(m)\ad(n)}(\phi)&= \Delta^{(m)\bd_m}_{~(\a_{m}}\cdots \Delta^{(1)\bd_1}_{~\a_{1})}\Delta^{(1)\b_1}_{~(\bd_{1}}\cdots\Delta^{(m)\b_m}_{~\bd_{m}}\phi_{\b(m)\ad_1\dots\ad_n)}~.
\end{align}
\end{subequations}
This makes it clear that both $\Pi^{(m,n)}$ and $\widehat{\Pi}^{(m,n)}$ act as the identity operator on the space of transverse fields of rank $(m,n)$.  


\subsection{Conformal higher-spin models}

As an application of our analysis, we would like to discuss the action for a conformal higher-spin field $\phi_{\a(m)\ad(n)}$ in AdS.\footnote{Gauge-invariant actions for conformal  higher-spin fields on arbitrary conformally flat $d=4$ backgrounds were constructed in \cite{KP2}. A few years earlier, Metsaev \cite{Metsaev2014}  developed the so-called ordinary derivative formulation for conformal higher-spin fields on AdS${}_d$.}
Making use of the linearised higher-spin Bach operators $\mathbb{B}^{(m,n)}, \widehat{\mathbb{B}}^{(m,n)}:V_{(m,n)}\rightarrow V_{(n,m)}$ defined by 
\begin{subequations}\label{Bach}
\begin{align}
\mathbb{B}_{\a(n)\ad(m)}(\phi)&=\mathcal{D}_{(\ad_1}{}^{\b_1}\cdots\mathcal{D}_{\ad_m)}{}^{\b_m}\mathcal{D}_{(\a_1}{}^{\bd_1}\cdots\mathcal{D}_{\a_n}{}^{\bd_n}\phi_{\b_1\dots\b_m)\bd(n)}~, \label{Bach1}\\
\widehat{\mathbb{B}}_{\a(n)\ad(m)}(\phi)&=\mathcal{D}_{(\a_1}{}^{\bd_1}\cdots\mathcal{D}_{\a_n)}{}^{\bd_n}\mathcal{D}_{(\ad_1}{}^{\b_1}\cdots\mathcal{D}_{\ad_m}{}^{\b_m}\phi_{\b(m)\bd_1\dots\bd_n)}~, \label{Bach2}
\end{align}
\end{subequations}
which were introduced in \cite{KP2}, the action takes the form
\begin{subequations}
\begin{align}
S_{\text{CHS}}^{(m,n)}[\phi, \bar{\phi}]&=\text{i}^{m+n}\int \text{d}^4x \, e \, \bar{\phi}^{\a(n)\ad(m)}\mathbb{B}_{\a(n)\ad(m)}(\phi)+{\rm c.c.}  \label{3.13}\\
&=\text{i}^{m+n}\int \text{d}^4x \, e \,\phi^{\a(m)\ad(n)}\widehat{\mathbb{B}}_{\a(m)\ad(n)}(\bar{\phi})+{\rm c.c.}  
\end{align}
\end{subequations}
This action is invariant under the gauge transformations 
\begin{align}
\delta_{\zeta}\phi_{\a(m)\ad(n)}=\mathcal{D}_{(\a_1(\ad_1}\zeta_{\a_2\dots\a_m)\ad_2\dots\ad_n)} 
~,
\label{3.99}
\end{align}
which is equivalent to the fact that the descendants \eqref{Bach} are gauge invariant.
The equation of motion for $\bar \phi_{\a(n)\ad(m)}$
 is the vanishing of the  higher-spin Bach tensor
\begin{align}
\mathbb{B}_{\a(n)\ad(m)}(\phi)=0~. \label{3.14}
\end{align}
The gauge freedom \eqref{3.99} allows us to impose the transverse gauge
\begin{align}
\phi_{\a(m)\ad(n)}\equiv \phi^{\text{T}}_{\a(m)\ad(n)}~,\qquad \mathcal{D}^{\g\gd}\phi^{\text{T}}_{\g\a(m-1)\gd\ad(n-1)}=0~. \label{gf}
\end{align}

There are three separate scenarios that we should consider, the first of which occurs when $m=n=s$. In this case the bosonic higher-spin Bach operator $\mathbb{B}^{(s,s)}$ coincides with $\mathbb{P}^{(s,s)}$,
\begin{align}
\mathbb{B}_{\a(s)\ad(s)}(\phi)=\mathbb{P}_{\a(s)\ad(s)}(\phi)~.
\end{align}
As a consequence of \eqref{3.10a}, this means that in the gauge \eqref{gf} the Bach tensor factorises,
\begin{align}
\mathbb{B}_{\a(s)\ad(s)}(\phi^{\text{T}})=\prod_{t=1}^{s}(\mathcal{Q}-\lambda_{(t,s,s)}\mathcal{S}^2)\phi^{\text{T}}_{\a(s)\ad(s)} \label{factor1}
\end{align}
and hence so too does the gauge fixed action. 

Next, let us consider the case when $n > m$. By taking appropriate derivatives of the Bach tensor \eqref{Bach1} one arrives at the following relation\footnote{Due to \eqref{2.1}, the left hand side of \eqref{3.19} is automatically totally symmetric in its dotted indices.}
\begin{align}
\mathcal{D}_{\ad_{m+1}}{}^{\a_{m+1}}\cdots\mathcal{D}_{\ad_{n}}{}^{\a_n}\mathbb{B}_{\a(n)\ad(m)}(\phi)=\mathbb{P}_{\a(m)\ad(n)}(\phi)~, \label{3.19}
\end{align}
which may be inverted to give
\begin{align}
\mathbb{B}_{\a(n)\ad(m)}(\phi)=\bigg[\prod_{t=m+1}^{n}\big(\mathcal{Q}-\lambda_{(t,m,n)}\mathcal{S}^2\big)\bigg]^{-1}\mathcal{D}_{\a_{m+1}}{}^{\bd_{1}}\cdots\mathcal{D}_{\a_{n}}{}^{\bd_{n-m}}\mathbb{P}_{\a(m)\ad(m)\bd(n-m)}(\phi)~.
\end{align}
It follows from \eqref{3.10a} that in the gauge \eqref{gf}, the Bach operator factorises as
\begin{align}
\mathbb{B}_{\a(n)\ad(m)}(\phi^{\text{T}})=\prod_{t=1}^{m}\big(\mathcal{Q}-\lambda_{(t,m,n)}\mathcal{S}^2\big)\mathcal{D}_{\a_{1}}{}^{\bd_{1}}\cdots\mathcal{D}_{\a_{n-m}}{}^{\bd_{n-m}}\phi^{\text{T}}_{\a_{n-m+1}\dots\a_{n}\ad(m)\bd(n-m)}~. \label{factor2}
\end{align}
We see that due to the mismatch of $m$ and $n$, the conformal operator $\mathbb{B}^{(m,n)}$ does not factorise wholly into products of second-order operators. However, using \eqref{3.19} it is easy to see that for transverse fields the following  equation can be derived from \eqref{3.14},
\begin{align}
\prod_{t=1}^{n}(\mathcal{Q}-\lambda_{(t,m,n)}\mathcal{S}^2)\phi^{\text{T}}_{\a(m)\ad(n)}=0~. \label{3.18}
\end{align}

If on the other hand $m > n$, then the Bach tensor may be written in terms of $\mathbb{P}^{(m,n)}$ 
\begin{align}
\mathbb{B}_{\a(n)\ad(m)}(\phi)=\mathcal{D}_{\ad_m}{}^{\b_1}\cdots\mathcal{D}_{\ad_{n+1}}{}^{\b_{m-n}}\mathbb{P}_{\a(n)\b(m-n)\ad(n)}(\phi)~.
\end{align}
Once again, from \eqref{3.10a} it follows that in the transverse gauge $\mathbb{B}^{(m,n)}$ factorises as
\begin{align}
\mathbb{B}_{\a(n)\ad(m)}(\phi^{\text{T}})=\prod_{t=1}^{n}\big(\mathcal{Q}-\lambda_{(t,m,n)}\mathcal{S}^2\big)\mathcal{D}_{\ad_{1}}{}^{\b_{1}}\cdots\mathcal{D}_{\ad_{m-n}}{}^{\b_{m-n}}\phi^{\text{T}}_{\a(n)\b(m-n)\ad_{m-n+1}\dots\ad_{m}}~. \label{factor3}
\end{align}
This time the higher-derivative equation derivable from \eqref{3.14} is 
\begin{align}
\prod_{t=1}^{m}(\mathcal{Q}-\lambda_{(t,m,n)}\mathcal{S}^2)\phi^{\text{T}}_{\a(m)\ad(n)}=0~. \label{3.20}
\end{align}
An interesting observation is that according to the definition \eqref{3.5}, when $m\neq n$ there is a discrete set of mass values corresponding to the range $\text{min}(m,n)<t\leq \text{max}(m,n)$ which are not partially massless but which enter the spectrum of the operators \eqref{3.18} and \eqref{3.20}.

To obtain the effective actions corresponding to the CHS models with $m\neq n$,
it is convenient to make use of the method of squaring which is always applied in the spinor theory.
In this method we have to  deal with the operator
\begin{align}
\widehat{\mathbb{B}}^{(n,m)}\mathbb{B}^{(m,n)}\phi_{\a(m)\ad(n)}=\prod_{t=1}^{m}\big(\mathcal{Q}-\lambda_{(t,m,n)}\mathcal{S}^2\big)\mathbb{P}^{(m,n)}\phi_{\a(m)\ad(n)}
\end{align}
which for a transverse field becomes
\begin{align}
 \widehat{\mathbb{B}}^{(n,m)}\mathbb{B}^{(m,n)}\phi^{\text{T}}_{\a(m)\ad(n)}=\prod_{t=1}^{m}\big(\mathcal{Q}-\lambda_{(t,m,n)}\mathcal{S}^2\big)\prod_{k=1}^{n}\big(\mathcal{Q}-\lambda_{(k,m,n)}\mathcal{S}^2\big)\phi^{\text{T}}_{\a(m)\ad(n)}~. 
 \end{align}

Finally, we note that in the case of unconstrained fields, the action \eqref{3.13} may be rewritten in terms of the projectors.  In the  ordinary bosonic $m=n=s$ 
case it takes the form
\begin{align}
S_{\text{CHS}}^{(s,s)}&=(-1)^s\int \text{d}^4x \, e \, \phi^{\a(s)\ad(s)}\prod_{t=1}^{s}(\mathcal{Q}-\lambda_{(t,s,s)}\mathcal{S}^2)\Pi_{\a(s)\ad(s)}(\phi)+{\rm c.c.}~,
\end{align}
while in the fermionic $m=n-1=s$ case it becomes
\begin{align}
S_{\text{CHS}}^{(s,s+1)}&=(-1)^s\text{i}\int \text{d}^4x \, e \, \bar{\phi}^{\a(s+1)\ad(s)}\prod_{t=1}^{s}(\mathcal{Q}-\lambda_{(t,s,s+1)}\mathcal{S}^2)\mathcal{D}_{\a_{s+1}}{}^{\bd}\Pi_{\a(s)\ad(s)\bd}(\phi)+{\rm c.c.}
\end{align}

For lower-spin values $s=3/2 $ and $s= 2,$ corresponding to $m=n-1=1$ and $m=n=2$ respectively, the factorisation of the conformal operators \eqref{Bach} was observed long ago in \cite{DeserN1,Tseytlin5, Tseytlin6}. 
This factorisation was conjectured, based on lower-spin examples, by Tseytlin
\cite{Tseytlin13} for the bosonic $m=n$ and fermionic $m=n-1 $ cases, 
and also by Joung and Mkrtchyan  \cite{Karapet1,Karapet2} 
for certain bosonic CHS models.
More recently, the factorisation was proved by several groups 
 \cite{ Metsaev2014, NT,GH} for those bosonic CHS models
on AdS${}_d$, with even $d$, 
 which are described by completely symmetric arbitrary spin conformal fields 
(the $m=n$ case in four dimensions). 

Using the formalism of spin projector operators in AdS${}_4$  developed in this work, 
the known factorisation properties for $m=n$ follow  immediately, 
 and are captured through the expression \eqref{factor1}. We have also provided the first derivation of the factorisation for conformal operators of arbitrary Lorentz type $(m/2,n/2)$, which is encapsulated by expressions \eqref{factor2} and \eqref{factor3}. This encompasses the case of arbitrary fermionic spin, $m=n-1$, 
 which to our knowledge was not covered previously in the literature. As in the bosonic case, we find that the spectrum of \eqref{factor2} and \eqref{factor3} consists of all partial masses. In contrast however, we find that the spectrum of the wave equations \eqref{3.18} and \eqref{3.20} contain a discrete set of massive values.

It would be interesting to re-derive our results using the ambient space approach, see \cite{BM, BG1, BG2} and references therein.


 \noindent
{\bf Acknowledgements:}\\
We are grateful to the organisers of the APCTP Workshop 
``Higher Spin Gravity: Chaotic, Conformal and Algebraic Aspects''
(Pohang, South Korea), 
where part of this work was completed, for the fantastic scientific atmosphere and generous support.
The work of SMK is supported in part by the Australian 
Research Council, project No. DP160103633.
The work of MP is supported by the Hackett Postgraduate Scholarship UWA,
under the Australian Government Research Training Program. 

 
\appendix
\section{Technical results}

Here we prove that the system of equations \eqref{eom} is invariant under depth $t$ gauge transformations \eqref{3.6} for the mass values \eqref{3.5}.

To begin with, it is clear that \eqref{eom1} is gauge invariant only if the gauge parameter is also on-shell with the same (pseudo-)mass $\mu$
\begin{align}
\big(\mathcal{Q}-\mu^2\big)\zeta_{\a(m-t)\ad(n-t)}=0~. \label{A.1}
\end{align}
Additionally, we need to ensure that the gauge variation of the transverse condition \eqref{eom2} vanishes,
\begin{align}
0=\mathcal{D}^{\b\bd}\delta_{\zeta}\phi_{\a(m-1)\b\ad(n-1)\bd}~. \label{A.2}
\end{align}
To compute the right hand side of \eqref{A.2} it is useful to 
 introduce the auxiliary commuting spinor variables $\U^{\a}$ and $\bar{\U}^{\ad}$. 
Associated with a tensor field $\phi_{\a(m)\ad(n)}$
of Lorentz type $(\frac{m}{2},\frac{n}{2})$
 is  a homogeneous polynomial $\phi_{(m,n)}(\U, \bar \U)$
of degree $(m,n)$ 
  defined by 
  \bea
  \phi_{(m,n)}:=\U^{\a_1}\cdots\U^{\a_m}\bar{\U}^{\ad_1}\cdots\bar{\U}^{\ad_n}\phi_{\a_1\dots\a_m\ad_1\dots\ad_n}~.
  \eea
  We denote the space of such homogeneous polynomials as $\mathcal{H}_{(m,n)}$.

Next we introduce two operators 
\begin{align}
\mathcal{D}_{(1,1)}:= \U^{\a}\bar{\U}^{\ad}\mathcal{D}_{\a\ad}~,\qquad \mathcal{D}_{(-1,-1)}:=\mathcal{D}^{\a\ad}\frac{\partial}{\partial\U^{\a}}\frac{\partial}{\partial \bar{\U}^{\ad}}~\equiv \mathcal{D}^{\a\ad}\partial_{\a}\bar{\partial}_{\ad}~,\label{A.4}
\end{align}
which increase and decrease
the degree of homogeneity by $(1,1)$ and $(-1,-1)$ respectively.
They may be shown to satisfy the algebra
\begin{align}
\big[\mathcal{D}_{(1,1)},\mathcal{D}_{(-1,-1)}\big]=(\bm\U+1)(\Box+4\mathcal{S}^2\bar{\bm M})+(\bar{\bm \U}+1)(\Box+4\mathcal{S}^2\bm M)~, \label{A.5}
\end{align}
where we have defined\footnote{The auxiliary variables $\U^{\a}$ and $\bar{\U}^{\ad}$ are inert with respect to the Lorentz generators.}
\begin{subequations}
\begin{align}
\bm\U&=\U^{\a}\partial_{\a}~,\qquad \qquad  \qquad~ \bm\U\phi_{(m,n)}=m\phi_{(m,n)}~,\\
\bar{\bm \U}&=\bar{\U}^{\ad}\bar{\partial}_{\ad}~,\qquad\qquad \qquad~\bar{\bm \U}\phi_{(m,n)}=n\phi_{(m,n)}~,\\
\bm M&=\U^{\a}\partial^{\b}M_{\a\b}~,\qquad\qquad \bm M\phi_{(m,n)}=-\frac{1}{2}m(m+2)\phi_{(m,n)}~,\\
\bar{\bm M}&=\bar{\U}^{\ad}\bar{\partial}^{\bd}\bar{M}_{\ad\bd}~,\qquad\qquad \bar{\bm M}\phi_{(m,n)}=-\frac{1}{2}n(n+2)\phi_{(m,n)}~.\label{A.6}
\end{align}
\end{subequations}
Then, via induction on $k$ it is possible to show that for any $\phi_{(m,n)}\in \mathcal{H}_{(m,n)}$, the following identity holds true
\begin{align}
&\big[\mathcal{D}_{(-1,-1)},\underbrace{\mathcal{D}_{(1,1)}\cdots\mathcal{D}_{(1,1)}}_{k\text{-times}}\big]\phi_{(m,n)}=-k(m+n+k+1)\big(\mathcal{Q}-\lambda_{(k,m+k,n+k)}\mathcal{S}^2\big) \notag\\
&\phantom{\big[\mathcal{D}_{(-1,-1)},\underbrace{\mathcal{D}_{(1,1)}\cdots\mathcal{D}_{(1,1)}}_{k\text{-times}}\big]\phi_{(m,n)}=}\times\underbrace{\mathcal{D}_{(1,1)}\cdots\mathcal{D}_{(1,1)}}_{(k-1)\text{-times}}\phi_{(m,n)}~.\label{A.7}
\end{align}

 Finally, using the operators \eqref{A.4}, one may show that the condition \eqref{A.2} is equivalent to
\begin{align}
0&=\mathcal{D}_{(-1,-1)}\underbrace{\mathcal{D}_{(1,1)}\cdots\mathcal{D}_{(1,1)}}_{t\text{-times}}\zeta_{(m-t,n-t)} \notag\\
 &=\underbrace{\mathcal{D}_{(1,1)}\cdots\mathcal{D}_{(1,1)}}_{t\text{-times}}\mathcal{D}_{(-1,-1)}\zeta_{(m-t,n-t)} \notag\\
 &\phantom{\mathcal{D}_{(1,1)}\cdots\mathcal{D}_{(1,1)}}-t(m+n-t+1)\big(\mathcal{Q}-\lambda_{(t,m,n)}\mathcal{S}^2\big)\underbrace{\mathcal{D}_{(1,1)}\cdots\mathcal{D}_{(1,1)}}_{(t-1)\text{-times}}\zeta_{(m-t,n-t)} \notag
\end{align}
where $t$ is the depth and we have used \eqref{A.7} in the second line. The first term vanishes if $\zeta_{\a(m-t)\ad(n-t)}$ is transverse
\begin{align}
0=\mathcal{D}^{\b\bd}\zeta_{\a(m-t-1)\b\ad(n-t-1)\bd}~,
\end{align}
whilst the second term vanishes if the mass in \eqref{A.1} satisfies $\mu^2=\lambda_{(t,m,n)}\mathcal{S}^2$.
\begin{footnotesize}

\end{footnotesize}

\end{document}

\end{document}